\documentclass[aps,prl,reprint,showpacs,showkeys]{revtex4-1}
\usepackage{microtype}
\usepackage{amsthm}
\usepackage{amsmath}
\usepackage{amsfonts}
\usepackage[mathscr]{eucal}
\usepackage{amssymb,epsfig,setspace}
\usepackage{graphicx}
\theoremstyle{remark}
\newtheorem{remark}{Remark}
\usepackage{natbib}
\usepackage[usenames]{xcolor}

\newcommand\scH{{\mathscr H}}

\newcommand\scO{{\mathscr O}}

\newcommand\mvector{\boldsymbol}

\newcommand\vA{\mvector{A}}

\newcommand\field{\mathbb}

\newcommand\C{\field{C}}

\renewcommand\Re{\operatorname{Re}}
\renewcommand\Im{\operatorname{Im}}

\newcommand\id{\operatorname{\mathrm{Id}}}

\newcommand\Dy{\frac{\mathrm{d}\phantom{y} }{\mathrm{d}y}}

\begin{document}

\title[Spectra calculations in the Bargmann representation]{Analytical method of spectra calculations in the Bargmann representation}%
\author{Andrzej J.~Maciejewski} \email{maciejka@astro.ia.uz.zgora.pl}
\affiliation{J.~Kepler Institute of Astronomy, University of Zielona
  G\'ora, Licealna 9, PL-65--417 Zielona G\'ora, Poland.}%
\author{Maria Przybylska}%
\email{M.Przybylska@if.uz.zgora.pl} \affiliation{ Institute of
  Physics, University of Zielona G\'ora, Licealna 9, 65--417 Zielona
  G\'ora, Poland }%
\author{Tomasz Stachowiak} \email{stachowiak@cft.edu.pl}
\affiliation{%
  Center for Theoretical Physics PAS, Al. Lotnikow 32/46, 02-668
  Warsaw, Poland }%

\date{\today}%
\begin{abstract}
  We formulate a universal method for solving an arbitrary quantum
  system which, in the Bargmann representation, is described by a
  system of linear equations with one independent variable, such
  as one- and multi-photon Rabi models, or $N$ level systems
  interacting with a single mode of the electromagnetic field and
  their various generalizations.  We explain three types of conditions
  that determine the spectrum and show their usage for two
  deformations of the Rabi model. We prove that the spectra of both
  models are just zeros of transcendental functions, which in one case
  are given explicitly in terms of confluent Heun functions.
\end{abstract}
                      
\maketitle
%
%
\graphicspath{{./Graphics/}}
\section{Introduction}

Our  goal and result is a general method which allows to properly
determine eigenvalues and eigenfunctions for a wide class of quantum
systems.  It is adequate for quantum optical setting where the
Bargmann representation allows for natural parametrization of the
electromagnetic degree of freedom and the resulting differential
equations are ordinary and linear. We then show its application to two
systems, which are generalizations of the famous Rabi model
characterized by the Hamiltonian
\begin{equation}
  \label{eq:HRabi}
  H = a^{\dag}a + \mu\sigma_z + 
  \lambda\sigma_x(a^{\dag}+a),
\end{equation}
where $a$, $a^{\dag}$ are the photon annihilation and creation
operators, $\mu$, $\lambda$ are the level separation and photon-atom
coupling constant, and $\sigma_x$, $\sigma_z$ are the Pauli spin
matrices.

This fundamental system describes interaction of a two-level atom with
a single harmonic mode of the electromagnetic field. Originally,
it was introduced to describe the effect of a rapidly varying, weak
magnetic field on an oriented atom possessing nuclear spin
\cite{Rabi:36::}.  It has been recently applied to a great variety
of physical systems, including cavity and circuit quantum
electrodynamics, quantum dots, polaronic physics and trapped ions, see
\cite{Gunter:09::,Niemczyk:10::,FornDiaz:10::,Crespi:12::,Casanova:10::,Braak:11::}.

Usually coupling between ``natural'' two-level atoms and the single
bosonic mode of radiation is quite weak and the rotating wave
approximation is valid. It leads to a solvable, the so-called
Jaynes-Cummings, model. However, recent achievements in circuit
quantum electrodynamics have enabled the exploration of such regimes,
e.g., the ultra-strong and the deep strong coupling regimes
of light-atom interaction so that the Jaynes-Cummings model begins to
fail. Effects of counter-rotating terms cannot be more neglected and
terms containing simultaneous excitation or deexcitation of both the
atom and the field must be taken into account
\cite{Gunter:09::,Niemczyk:10::,FornDiaz:10::,Crespi:12::,Casanova:10::}.
The second reason for its recent renaissance is the realization that
the strong coupling regimes might require more interaction terms than
just those mentioned above. One such generalization, the so-called
Rabi model with broken symmetry, was proposed in \cite{Braak:11::} and
its additional term was justified physically as spontaneous emission
by the atom. The Hamiltonian of this generalization is 
\begin{equation}
  \label{eq:Heps}
  H_{\varepsilon} = a^{\dag}a + \mu\sigma_z + 
  \lambda\sigma_x(a^{\dag}+a) +\varepsilon \sigma_x.
\end{equation}
This will be the first example we study. The second one was proposed
in \cite{Grimsmo:13::a,Grimsmo:14::a}. It includes a nonlinear
coupling term between the atom and the cavity:
\begin{equation}
  H= \left(\omega +\dfrac{U}{2}\sigma_z\right) a^{\dag}a +
  \dfrac{\omega_0}{2}\sigma_z + 
  g\sigma_x(a^{\dag}+a).
  \label{HNL}
\end{equation}
An alternative physical motivation of the additional term is that
it could arise in the dispersive limit of the Jaynes-Cummings
model. However, the first possibility is more accessible
experimentally as described in \cite{Grimsmo:13::a}. We chose to keep
the notation of that paper, for the second model, to facilitate
comparison. A quick inspection shows that the parameters of models
\eqref{eq:Heps} and \eqref{HNL} are related via
\begin{equation}
  \omega = 1,\quad \omega_0 = 2\mu,\quad g=\lambda.
\end{equation}

Although the spectrum of the classical Rabi model has been determined
by numerical and approximate methods before, see, e.g.,
\cite{Feranchuk:96::,Durst:86::,Tur:01::,Emary:02::a}, there still
is a lack of a general approach which works well for arbitrary
parameters values and which has a solid mathematical
foundations. Recently several approaches devoted to determination
of the spectrum of this and similar models have appeared, see, e.g.,
\cite{Braak:11::,Braak:13::a,Braak:13::d,Moroz:12::a,Chen:12::}, and
references therein.  The authors have also applied the present method
as outlined in the preliminary preprint \cite{Maciejewski:12::x}, to
determine the full spectrum in \cite{Maciejewski:14::}, including some
isolated points that are usually overlooked.

It should be underlined that the Rabi model is one of the simplest
ones in quantum physics. This is why the knowledge of its exact
eigenvalues and eigenfunctions is of great theoretical importance.
Although the question about the spectrum and eigenstates comes from
physics, it is a mathematical one. It is obvious that unjustified
methods may lead to incorrect physical interpretations of considered
models.

In the Bargmann-Fock representation, see~\cite{Bargmann:61::}, the two-component wave function $\psi=(\psi_1, \psi_2)$ is an element
of Hilbert space $\mathscr{H}^2=\mathscr{H}\times \mathscr{H}$, where
$\mathscr{H}$ is the Bargmann-Fock Hilbert space of entire functions
of one variable $z\in\mathbb{C}$. The elegant connection with the
standard picture is that the annihilation and creation operators $a$,
and $a^{\dag}$ become $\partial_z$ and multiplication by $z$,
respectively, for clearly $[\partial_z,z]=1$. The scalar product
in $\mathscr{H}$ is given by
\[
\langle
f,g\rangle=\dfrac{1}{\pi}\int_\mathbb{C}\overline{f(z)}g(z)e^{-|z|^2}\mathrm{d}
(\Re(z))\mathrm{d}(\Im(z)).
\]
It is worth mentioning that this space was also introduced,
independently of Bargmann, by J. Newman and H. S. Shapiro
\cite{Newman:64::,Newman:66::}. However their motivation was connected
with works of Ernst Fischer \cite{Fischer:11::,Fischer:17::}.  They tried
to generalize a very beautiful construction of E.~Fisher valid for
polynomials.

The Hilbert space $\mathscr{H}$ has several peculiar properties. Let
us mention two of them:
\begin{enumerate}
\item $f(z)\in\mathscr{H}$ does not imply that $f'(z)\in\mathscr{H}$.
\item $f(z)\in\mathscr{H}$ does not imply that $zf(z)\in\mathscr{H}$.
\end{enumerate}
To understand these rather strange properties we have to recall some
definitions and facts from the theory of entire functions, see
\cite{Levin:96::,Boas:54::}. If $f(z)$ is an entire function, then to
characterize its growth, the following function is used:
\begin{equation}
  M_f(r) := \underset{|z|=r}{\mathrm{max}}|f(z)|.
\end{equation}
We omit the subscript $f$ later on, because the investigated function
is known from the context. If for an entire function $f(z)$ we have
\begin{equation}
  \lim_{r\rightarrow\infty}\sup\frac{\ln(\ln M(r))}{\ln r} = \varrho,\quad
  \mathrm{with}\quad 0\leq \varrho \leq \infty,
\end{equation}
then $\varrho$ is called the order (or growth order) of $f(z)$. If,
further, the function has positive order $\varrho<\infty$ and
satisfies
\begin{equation}
  \lim_{r\rightarrow\infty}\sup\frac{\ln M(r)}{r^\varrho} =\sigma,
\end{equation}
then we say that $f(z)$ is of order $\varrho$ and of type $\sigma$.

Assume that $f(z)$ belongs to $\mathscr{H}$, then one can prove the
following facts \cite{Bargmann:61::}:
\begin{enumerate}
\item $f(z)$ is of order $\varrho\leq 2$.
\item If $\varrho=2$, then $f(z)$ is of type $\sigma\leq\tfrac12$.
\end{enumerate}
If $\varrho=2$ and $\sigma=\tfrac12$, then the question whether
$f(z)\in\mathscr{H}$ requires a separate investigation. Particularly in the
mentioned case when $f(z)\in\mathscr{H}$ but $f'(z)\notin\mathscr{H}$
the function is of order $\varrho=2$ and type $\sigma=\tfrac12$. For
additional details see \cite{Vourdas:06::}.

The usefulness of this representation can immediately be seen with the
harmonic oscillator, which represents the radiation. The
time-independent Schr{\"o}dinger equation for energy $E$ is simply
$H\psi(z) = z\psi'(z)=E\psi(z)$ and one immediately recovers the
orthonormal eigenbasis as $\{z^n/\sqrt{n!}\}_{n\in\mathbb{N}}$. The
connection with the usual space of square-integrable functions of $q$
is given by the integral kernel
$\exp\left(-(z^2+q^2)/2+\sqrt{2}qz\right)$ which is one of the forms
of the generating function for the Hermite polynomials. Each $z^n$
thus corresponds to the appropriately normalized wave function
$e^{-q^2/2}H_n(q)$. In this basis the operator $a$ is just an infinite
matrix with entries on the superdiagonal, so all the mentioned
Hamiltonians can be constructed as tensor products of such matrices
with the sigma matrices. This allows for direct numerical
diagonalization. However, the open question that we wish to tackle
is how to determine the spectrum rigorously with as explicit exact
formulas as possible.

In the Bargmann-Fock representation energy $E$ belongs to the spectrum
of the problem, if and only if, for this value of $E$ the equation
$H\psi=E\psi$ has entire solution $\psi= (\psi_1,
\psi_2)\in\mathscr{H}^2$.  We want to find, if possible, a formula for
those values of $E$.
 
As we already mentioned, in the Bargmann-Fock representation, the
considered models are described by a system of linear differential
equations. We shall see that the equations in question will involve
regular singular points and a possibly irregular point at infinity
on the complex $z$ plane. The conditions that the considered system
has a solution with components belonging to $\scH$, are roughly
threefold:
\begin{itemize}
\item Local conditions. At each regular singular point $z=s$ there
  exists at least one solution which is holomorphic on an open set
  containing $s$.
\item Global conditions.  Among all solutions which are locally
  holomorphic, we can find at least one at each singular point such
  that they are a holomorphic continuation of one another.
\item Normalization conditions. The entire function obtained in the
  above way must have finite Bargmann norm.
\end{itemize}
Our method gives straightforward and natural compatibility conditions
in term of Wronskian determinants and was first formulated in our
unpublished preprint \cite{Maciejewski:12::x}. For simplicity sake
we chose the two models that can be given either as a system of two
first order equations or one equation of the second order. The
application to higher order equations, as those investigated
in \cite{Braak:13::c} or \cite{Travenec:12::}, will appear in future
work \cite{Kus:14::}.

In the Bargmann representation, the first considered model
is described by the following system of two differential equations
\begin{equation}
  \begin{split}
    &(z+\lambda)\dfrac{\mathrm{d}\psi_1}{\mathrm{d}z}= (E-\varepsilon
    -\lambda
    z)\psi_1-\mu\psi_2,\\
    &(z-\lambda)\dfrac{\mathrm{d}\psi_2}{\mathrm{d}z}=
    (E+\varepsilon+\lambda z)\psi_2-\mu\psi_1.
  \end{split}
  \label{eq:syste}
\end{equation}
We will use this model to illustrate the single equation approach
below.  The second system takes the form
\begin{equation}
  \begin{split}
    \left(\omega+\dfrac{U}{2}\right)z\psi_1'+\dfrac{\omega_0}{2}\psi_1+g\psi_2'+gz\psi_2=E\psi_1,\\
    \left(\omega-\dfrac{U}{2}\right)z\psi_2'-\dfrac{\omega_0}{2}\psi_2+g\psi_1'+gz\psi_1=E\psi_2.
  \end{split}
  \label{eq:bargmanki}
\end{equation}
When we change the independent variable
\begin{equation}
  z\to y=\frac{\sqrt{4\omega^2-U^2}}{2g} z,
\end{equation}
then this system can be rewritten in the matrix form
\begin{equation}
  \Dy \psi =\vA(y)\psi,
  \label{eq:systmat} 
\end{equation} 
where matrix $\vA$ has the following entries
\begin{equation}
  \begin{split}
    a_{11}&=\dfrac{y (-4 g^2 + (U - 2 \omega) (2 E - \omega_0))}{(U^2
      - 4
      \omega^2)(y^2-1)},\\
    a_{12}&=-\dfrac{4g^2y^2+(U + 2 \omega) (2 E + \omega_0)}{(U + 2 \omega) \sqrt{-U^2 + 4 \omega^2}(y^2-1)},\\
    a_{21}&=\dfrac{4 g^2 y^2-(U - 2 \omega) (2 E - \omega_0)}{(U - 2 \omega) \sqrt{-U^2 + 4 \omega^2}(y^2-1)},\\
    a_{22}&=-\dfrac{y\left(4 g^2 + (U + 2 \omega) (2 e +
        \omega_0)\right)}{(U^2 - 4 \omega^2)(y^2-1)}.
  \end{split}
\end{equation}
In the above one must assume $4\omega^2\neq U^2$, and if that is not
the case the type and position of singularities changes
considerably. This special case requires the aforementioned closer
investigation of the behaviour at infinity which we will present
in our future work \cite{ams:2014::}.

Both our examples share the same features. They are described by
linear systems with rational coefficients and in both cases the
problem is to distinguish those parameters values for which the system
admits a solution in $\mathscr{H}$. As is well known, a system
of linear equations can be reduced to one equation of higher
order. For example, an elimination of $\psi_2(z)$
from~\eqref{eq:syste} gives one second order equation for
$\varphi(z):=\psi_1(z)$ of the form
\begin{equation}
  \label{eq:s}
  \varphi'' + p(z) \varphi' + q(z) \varphi=0, 
\end{equation}
with
\begin{equation}
  \label{eq:pq}
  \begin{split}
    p(z)&=- \dfrac{\lambda + 2 \epsilon \lambda +
      z ( 2 E -1 + 2 \lambda^2)}{z^2 - \lambda^2},\\
    q(z)&=-\dfrac{\epsilon^2-E^2 + 2 z \epsilon \lambda + \lambda
      (\lambda + z ( z \lambda -1)) + \mu^2}{z^2 - \lambda^2}.
  \end{split}
\end{equation}

\section{Method}

First, we make some general remarks. Although an arbitrary system
of $n$ linear differential equations with rational coefficients of the
first order can be transformed into a single linear equation, in some
cases it seems natural to work directly with the given
system. However, for a system to decide which singular point
is regular, and what the exponents are at this point is not so
obvious. All known algorithms which allow to determine these basic
characteristics of singular points for a system are rather
involved. Still, the general procedure described below can easily be
adapted to a system, only some technical points are more intricate \cite{Kohno:99::}.

We make the following assumptions. The considered equation has the
form
\begin{equation}
  \label{eq:s2}
  \varphi'' + p(z) \varphi' + q(z) \varphi=0, 
\end{equation}
where $p(z)$ and $q(z)$ are rational functions. All singular points
$\Sigma:=\{s_1,\ldots,s_m\}\subset\mathbb{C}$ are regular and they are poles of the equation's coefficients. Thus, all poles of $p(z)$ are of order not
grater than 1, and all poles of $q(z)$ are of order not greater than
2. The infinity is an irregular singular point (in all examples
of quantum optics we know it seems to be the case). Thus either
\begin{equation}
  P(\zeta) = \frac{2}{\zeta}-\frac{1}{\zeta^2}p\left(\frac{1}{\zeta}\right)
\end{equation}
has a pole of order greater than 1 at $\zeta=0$, or
\begin{equation}
  Q(\zeta) := \frac{1}{\zeta^4}q\left(\frac{1}{\zeta}\right)
\end{equation}
has a pole of order greater than 2 at $\zeta=0$.

We also assume that the singular points are numbered and located
in such a way that there exist open disks $D(s_i,r_i)$ centred
at $s_i$ and of radius $r_i$ satisfying the following conditions:
\begin{enumerate}
\item $\Sigma\cap D(s_i,r_i)=\{s_i\}$ for $1\leq i\leq m$.
\item $U_i : = D(s_i,r_i)\cap D(s_{i+1},r_{i+1})\neq\emptyset$ for
  $1\leq i <m$.
\end{enumerate}
\begin{remark}
  As a matter of fact our method works for an arbitrary configuration
  of singular points. However for such general configuration one has
  to use the continuation theory, see, e.g., \cite[Ch.~IX]{Conway:78::}.
\end{remark}
We can now return to the description of our method itself. Assume that
$\varphi(z)$ is an entire solution
of \eqref{eq:s2}. If $z_0\notin\Sigma$, then all solutions
of \eqref{eq:s2} are locally holomorphic in a neighbourhood
of $z_0$. So, local conditions do not give any restrictions on the
parameters. At a singular point $s_i\in\Sigma$ the form of local
solutions depends on exponents at this point. We denote them by
$\rho_i$ and $\varsigma_i$. A general basis of solutions is
\begin{equation}
  \begin{split}
    \phi_{i1} &= (z-s_i)^{\rho_i}h_{i}(z),\\
    \phi_{i2} &= l_{i}\phi_{i1}\ln(z-s_i) +
    (z-s_i)^{\varsigma_i}k_{i}(z),
  \end{split}
\end{equation}
where $h_i(z)$ and $k_i(z)$ do not vanish at $z=s_i$ and are locally
holomorphic with radius of convergence not smaller than the distance
from $s_i$ to the closest other singular point.  The logarithmic term
might arise only when the difference of exponents is an integer,
although this condition is not sufficient.  By assumption, the power
expansion of $\varphi(z)$ at $s_i$ is holomorphic.  This implies that
at least one exponent, let us say $\rho_i$ must be a non-negative
integer, i.e., $\rho_i\in\mathbb{N}=\{0,1,\ldots\}$ for all
$i$. In other words, $\rho_i$ is either the largest integer exponent
or the only integer exponent. Hence, at each singular point we have a
locally holomorphic solution $\varphi_i(z)$ of the form
\begin{equation}
  \varphi_i(z) = \zeta_i\, (z-s_i)^{\rho_i}h_i(z)+
  \xi_i\, (z-s_i)^{\varsigma_i}k_i(z),
  \label{fii}
\end{equation}
and $\xi_i$ is nonzero only when $\varsigma_i$ is an integer and
$l_i=0$ so that the logarithmic term vanishes. Note  that the
space of locally holomorphic solutions around $s_i$ could be at most 
two-dimensional. If that happens at a point $s_i$ with $1<i\leq m$, then a local 
holomorphic solution around  point $s_{i-1}$ can always be
decomposed as linear combination \eqref{fii} because $ \phi_{i1} $ and  $ \phi_{i2} $ is the  basis of solutions, so the constants $\zeta_i$ and $\xi_i$ are fixed. This way the
two local expansions $\varphi_{i-1}$ and $\varphi_i$ coincide on $U_{i}$ and there are no additional
conditions at this point.

Similarly, if there were two solutions at $s_{i}$, hence at all $s_j$
for $j<i$, but only one holomorphic solution at $s_{i+1}$, then this
single solution can always be decomposed into $\varphi_{i}$
as in \eqref{fii}. This fixes the constants $\zeta_i$ and $\xi_i$ and
we proceed with this solution to the next singular point. Note that this will also fix all the previous constants $\zeta_j$ and $\xi_j$ because we are left with only one solution.

Finally, if there is only a single solution $\varphi_i$ at $s_{i}$ and
likewise $\varphi_{i+1}$ at $s_{i+1}$, they must be linearly dependent
in order for them to be expansions of the same entire function. So,
there exist $\mathbb{C}\ni(\alpha_i,\beta_i)\neq(0,0)$, such that
\begin{equation}
  F_i(z) :=
  \alpha_i\varphi_i(z)+\beta_i\varphi_{i+1}(z)=0
  \quad \textrm{for all } z\in U_i.
  \label{E1}
\end{equation}
As $F_i(z)$ vanishes on a non-empty open set on which
it is holomorphic, it vanishes identically. Thus we also have
\begin{equation}
  F_i'(z) :=
  \alpha_i\varphi_i'(z)+\beta_i\varphi_{i+1}'(z)=0
  \quad \textrm{for all } z\in U_i.
  \label{E2}
\end{equation}
From \eqref{E1} and \eqref{E2} we deduce
\begin{equation*}
  \label{eq:w1}
  W_i(z):=\det\begin{bmatrix}
    \varphi_i(z) &  \varphi_{i+1}(z) \\ 
    \varphi_i'(z) &  \varphi_{i+1}'(z) 
  \end{bmatrix}=0, \quad \textrm{for all } z\in U_i.
\end{equation*}
But $W_i(z)$ is the Wronskian of two solutions of the same equation,
so if it vanishes at one point $z_i\in U_i$, then it vanishes
identically on $U_i$.  Proceeding this way we obtain at most 
 $m-1$ ``gluing'' conditions
$W_i(z_i)=0$ for $1\leq i<m$, which guarantee that $\varphi(z)$
is an entire function.

However, functions distinguished by these conditions are not
necessarily elements of $\mathscr{H}$. We also need to check
if $\langle \varphi,\varphi \rangle<\infty$. As mentioned above
it depends primarily on the order and type of the entire functions.
If infinity is a regular point, then necessarily
$\varphi(z)\in\mathscr{H}$, because the growth is subexponential.

In general, if the infinity is an irregular singular point the problem
is really hard especially when the considered equation is of order
higher than two. One has to determine  asymptotic
expansions of solutions at infinity. In the simplest case, which holds for the two
models considered here, they are of the form
\begin{equation}
  \varphi\sim e^{\sigma z^{\varrho}}z^{\rho}
  \left(1+\scO\left(\frac{1}{z}\right)\right).
\end{equation}
These series are formal, but they give bounds for the growth order $\varrho$
of the function, and its type $\sigma$ \cite{Kohno:99::}. 

To investigate these problems in  whole generality quite involved 
mathematical techniques must be used. 
So, the whole  exposition   of  more general version
of our method  will be published separately.

\section{Application to the first model}
Our method applied to the generalized Rabi model~\eqref{eq:syste}
provides a closed-form formula for the spectrum of the problem.
It is given as zeros of a certain transcendental function $W(p)$
expressed in terms of the confluent Heun functions. Here
$p:=(x,\lambda, \mu, \epsilon)$, $x:=E+\lambda^2$ is taken as a
spectral parameter.  System~\eqref{eq:syste} written as the second
order equation~\eqref{eq:s} after change of variables
\begin{equation*}
  \label{eq:vy}
  v(y):=\exp(-2\lambda^2y)\psi_1(\lambda(2y-1)), \qquad z=\lambda(2y-1),
\end{equation*}
transforms to the confluent Heun equation
\begin{equation}
  v''+\left(\alpha+\dfrac{\beta+1}{y}+\dfrac{\gamma+1}{y-1}
  \right)v'+\left(\dfrac{ \widetilde\mu }{
      y}+\dfrac{\widetilde\nu}{y-1}\right)v=0,
  \label{eq:Heun}
\end{equation} 
where
\begin{equation}
  \begin{split}
    \widetilde\mu&=\dfrac{1}{2}(\alpha-\beta-\gamma+\alpha\beta-\beta\gamma)-\eta,
    \\
    \widetilde\nu&=\dfrac{1}{2}
    (\alpha+\beta+\gamma+\alpha\gamma+\beta\gamma)+\delta+\eta.
  \end{split}
\end{equation}
In terms of $(x,\lambda,\mu,\epsilon)$ the above parameters are given
by
\begin{equation}
  \begin{split}
    &\alpha=4\lambda^2,\qquad \beta=-x+\epsilon,\\
    &\gamma=-1-x-\epsilon, \qquad\delta= 2(1-2\epsilon)\lambda^2,\\
    &2\eta= 1 - 2 \mu^2 + (1+x)(x - 4 \lambda^2 )+ \epsilon (1 + 4
    \lambda^2)-\epsilon^2.
  \end{split}
  \label{eq:par_heun_def}
\end{equation} 
The characteristic exponents at regular singularities $y=0$ and $y=1$ are $\{0,-\beta\}$ and $\{0,-\gamma\}$, respectively.

If $v_1(p;y)$ and $v_2(p;y)$ are two solutions of~\eqref{eq:Heun},
then their Wronskian is defined in usual way
\begin{equation}
  \label{eqw}
  w[v_1,v_2](p;y):= v_1'(p;y)v_2(p;y)-v_1(p;y)v_2'(p;y).
\end{equation}
But to simplify notation, to denote this Wronskian we just write
$w(p;y)$ if solutions $v_1$ and $v_2$ are specified.

Solutions of Eq.~\eqref{eq:Heun} are the confluent Heun
functions, see~\cite{Ronveaux:95::},
\begin{equation}
  \label{eq:hc}
  H_1(y):=\operatorname{HeunC}(a_0
  ; y), \quad  H_2(y):=\operatorname{HeunC}(a_1;1- y), 
\end{equation}
with parameter sets $a_0:=(\alpha,\beta,\gamma,\delta,\eta)$, and
$a_1:=(-\alpha,\gamma,\beta,-\delta,\delta+\eta)$.  The continuation
condition is given by their Wronskian
\begin{equation}
  \label{eq:w}
  w(p;y):= H_1(y)H_2'(y) - H_1'(y)H_2(y),
\end{equation}
in the following way $W(p):=w(p;1/2)=0$.

In the above formulae we assumed that neither $-\gamma=1+x+\epsilon$,
nor $-\beta=x-\epsilon$ is a non-negative integer. This is the generic
case, where only one characteristic exponent around each point
is integer (equal to zero). This only leaves out at most a finite
number of spectrum points, as described below.

\begin{figure}[t]
  \begin{center}
    \includegraphics[width=1\columnwidth,clip]{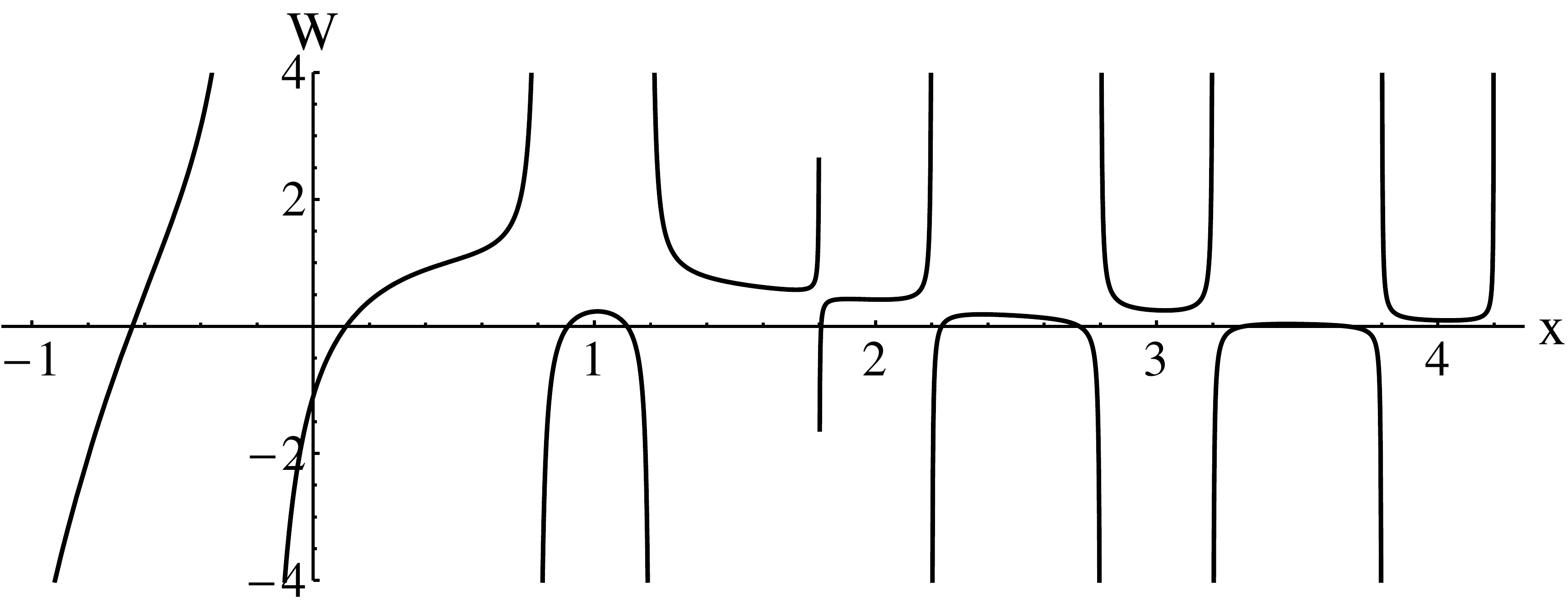}
  \end{center}
  \caption{\label{fig:3} Graph of Wronskian $W(p)$ for
    $p:=(x,\lambda,\mu,\epsilon)=(x,4/10,7/10, 1/5)$.}
\end{figure}

\begin{figure}[t]
  \begin{center}
    \includegraphics[width=0.9\columnwidth,clip]{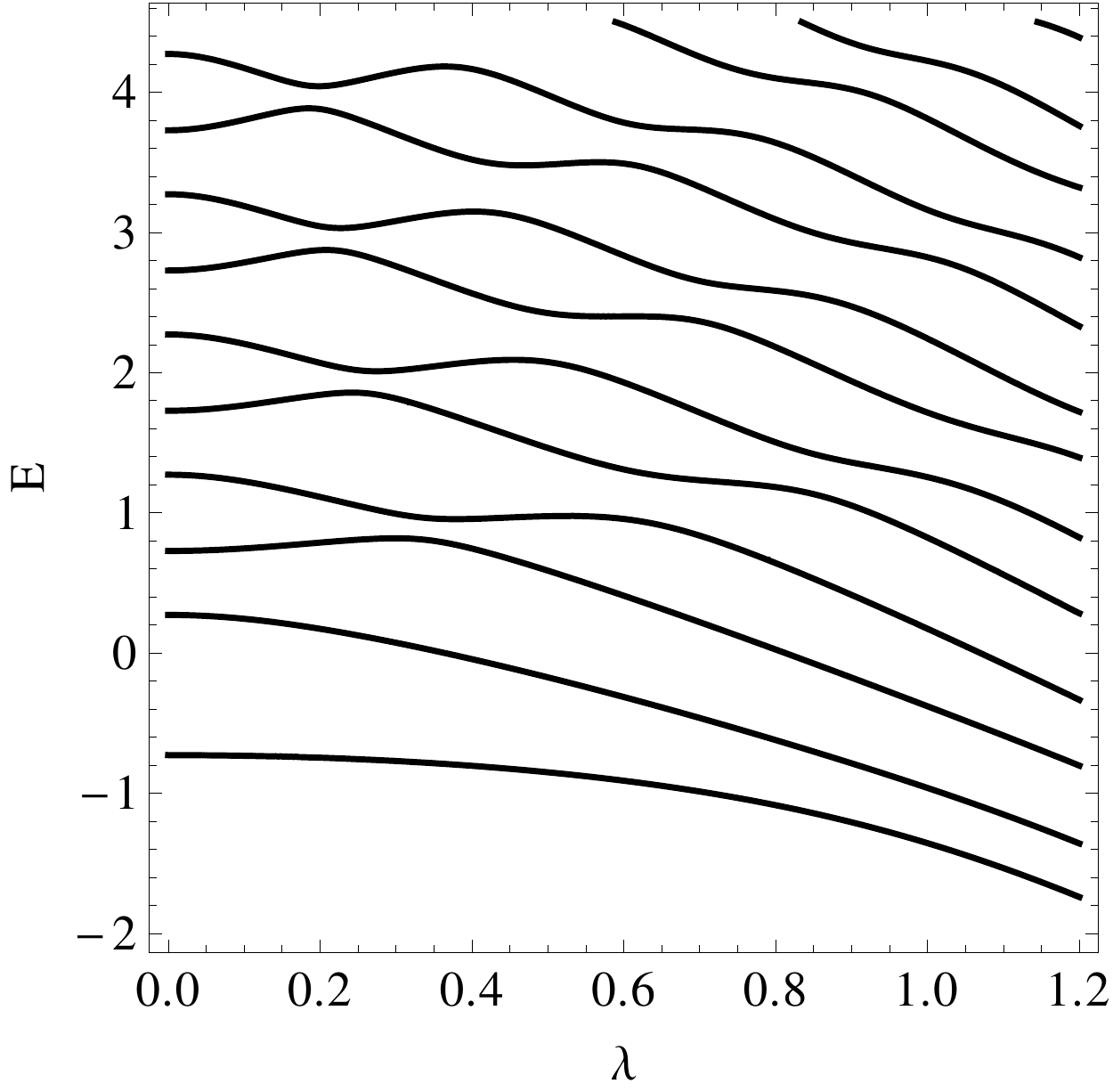}
  \end{center}
  \caption{\label{fig:5} Spectrum of generalized Rabi model for
    $\mu=0.7$, and $\epsilon=0.2$.}
\end{figure}

The graph of $W(p)$ and the spectrum for $\epsilon=0.2$ are shown
in Figs.~\ref{fig:3} and~\ref{fig:5}, respectively.

\begin{figure}[ht]
  \begin{center}
    \includegraphics[width=1\columnwidth,clip]{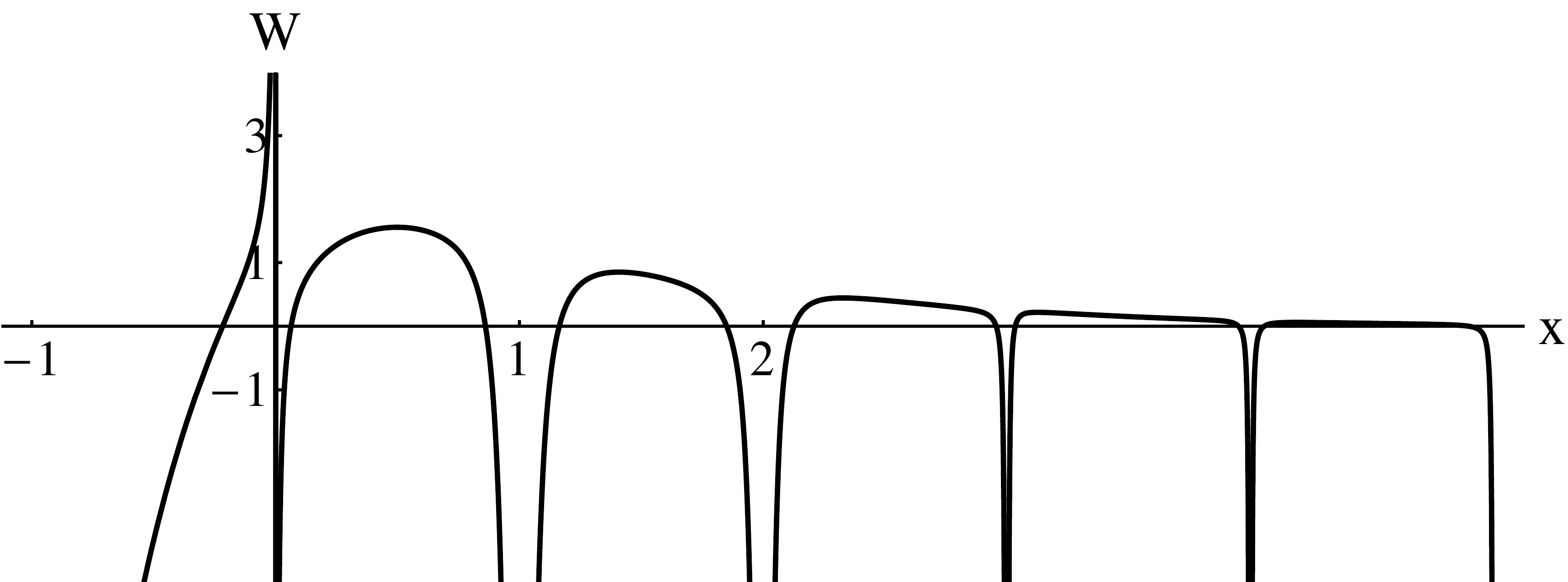}
  \end{center}
  \caption{\label{fig:4} Graph of Wronskian $W(p)$ for
    $p:=(E,\lambda,\mu,\epsilon)=(E,7/10, 4/10,0)$.}
\end{figure}

\begin{figure}[ht]
  \begin{center}
    \includegraphics[width=0.9\columnwidth,clip]{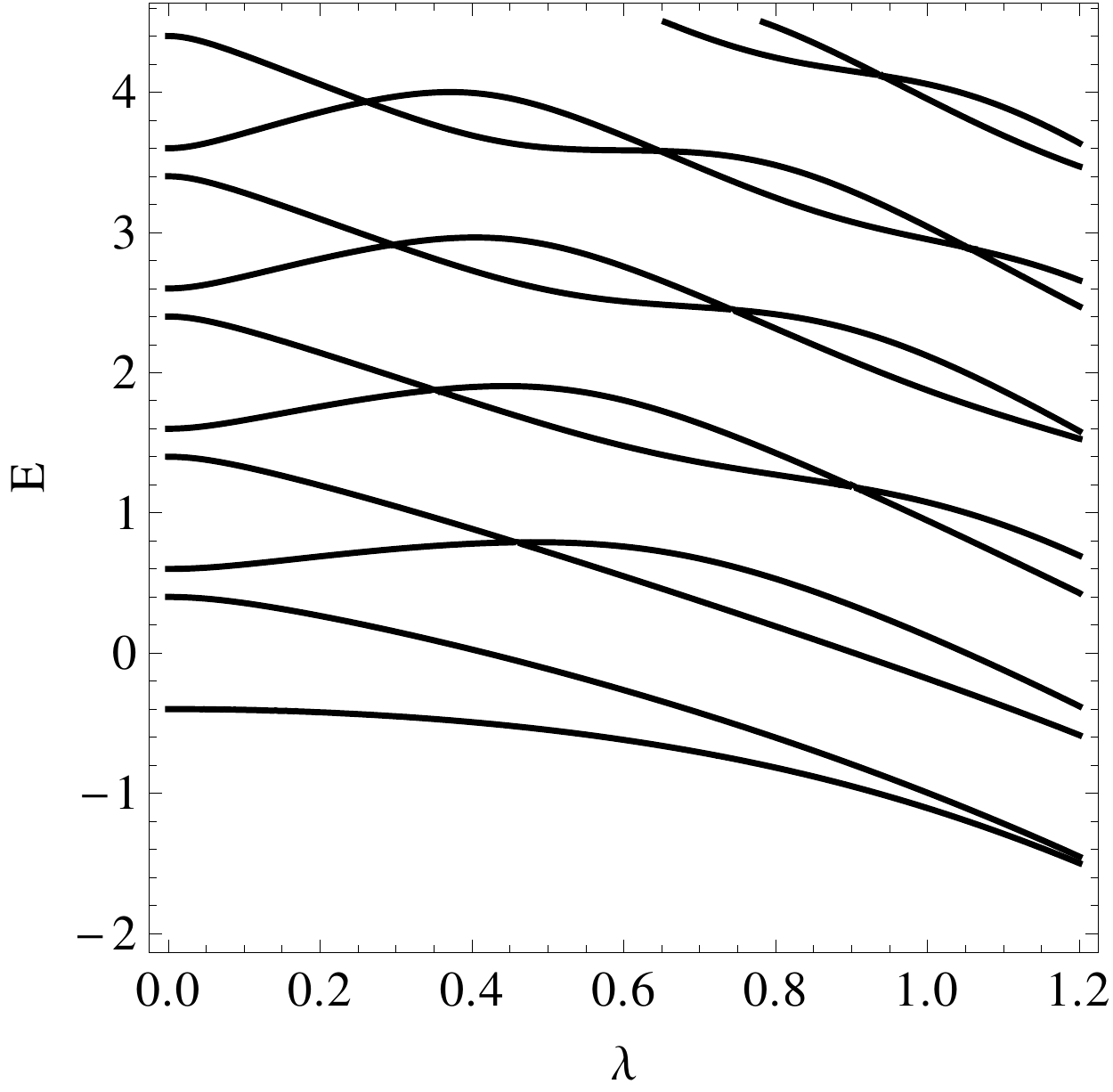}
    \caption{\label{fig:44} Spectrum of the Rabi model for $\mu=0.4$.
    }
  \end{center}
\end{figure}

For comparison, we show the graph of $W(p)$ as a function of $x$ for
the Rabi model (i.e., for $\epsilon=0$) in Fig.~\ref{fig:4}. The
corresponding spectrum is shown in Fig.~\ref{fig:44}.


When one of the other characteristic exponents is a natural number,
the Heun function might involve a logarithm but then there always
exists a local holomorphic solution corresponding to the larger
exponent. These solutions are given by
\begin{equation}
  \begin{split}
    H_3(y)&:=y^{-\beta}(y-1)^{-\gamma}\mathrm{HeunC}(c_0,y),\\
    H_4(y)&:=(y-1)^{-\gamma}\mathrm{HeunC}(c_1,1-y),
  \end{split}
\end{equation}
where $c_0:=(\alpha,-\beta,-\gamma,\delta,\eta)$ and
$c_1:=(-\alpha,-\gamma,\beta,-\delta,\delta+\eta)$. Additionally, the
logarithmic term can vanish, making both expansions at a given point
locally holomorphic. This happens when the parameters satisfy
an additional constraint which coincides with the $\Delta_n$ condition
by Fiziev, who showed that the confluent Heun function is then just a polynomial
\cite{Fiziev:10::}.  Since both of these options could arise at any
of the two points, let us introduce the following notation,
$\varsigma_s$  will
denote the exponent in question at a point $s\in \{0,1\}$, and
$\Delta_s:=\Delta_{\varsigma_s}(c_s)$. Keep in mind that the $\Delta$
condition depends on parameters differently at each point and also
that it is given recursively and the integer exponent $\varsigma_s$
specifies how far the recurrence is carried. We then end up with three
possible sub-cases. For each of them we have to specify two solutions
for calculations of Wronskian.
\begin{enumerate}
\item At one singular point $s\in\{0,1\}$ both exponents are integer and
  there are logarithmic therms in local solutions,
  i.e.  $\varsigma_s\in\mathbb{N}$ and $\Delta_s\neq 0$.  Moreover,
  at the other singular point only one exponent is an integer,
  i.e.  $\varsigma_{1-s}\neq\mathbb{N}$. In this case we take
  $H_{s+3}$ and $H_{2-s}$ as two solutions for calculation of the
  Wronskian.
\item If, for both $s\in\{0,1\}$, $\varsigma_s\in\mathbb{N}$ and
  $\Delta_s\neq 0$, then take $H_3$ and $H_4$. Notice that in this
  case, logarithmic terms in local solutions around each point appear.
\item If for one point $s\in\{0,1\}$ we have
  $\varsigma_s\in\mathbb{N}$, $\Delta_s=0$, then both local solutions
  around this point are holomorphic. One can show that,  in this case, 
  Heun equation~\eqref{eq:par_heun_def} has an entire solution so the
  continuation condition if fulfilled.
\end{enumerate}
The third subcase is rather remarkable because the vanishing of the
logarithm therm gives us two locally holomorphic solutions around one
point, and there is always at least one good solution around the
second point---either one corresponding to the zero exponent, or the
other corresponding to the second integer exponent. A locally
holomorphic solution around the second singular point must be a linear
combination of the two local solutions around the first singular
point, as stated in the Method section. As it can be continued throughout a set
containing both singular points, its radius of convergence is larger than the
distance between those points, so it must be infinite.

The sets of parameter values for each $\Delta$ condition can
intersect, meaning that there could be two integer exponents, which
in turn give two points of spectrum $x_1=\epsilon-\beta$ and
$x_2=-\epsilon-\gamma-1$, each with one entire
eigenstate. It is further possible that the $x_i$ coincide giving a
degenerate energy level with two entire eigenstates, and since both
exponents are integers it follows that in such a case necessarily
$2\epsilon\in\mathbb{Z}$. This is the most ``degenerate'' case, when
both logarithmic terms vanish. Additionally, if the exponents differ by one,
i.e. $\beta=\gamma+1$, the above implies $\epsilon=0$ and we recover
the classical Judd states of the unperturbed Rabi model described, e.g., in \cite{Kus:86::,Maciejewski:14::}.

Finally, concerning the finiteness of the norm, we observe that
for differential equation \eqref{eq:s} with $p(z)$ and $q(z)$ given
by~\eqref{eq:pq}, the Poincare rank is 1, and the asymptotic expansions of its solutions
are
\begin{equation}
    \begin{split}
        \varphi_1(z)&\sim e^{\lambda z}z^{E+\lambda^2+\epsilon-1}
        \left(1 +
        \scO\left(\frac{1}{z}\right) \right),\\
        \varphi_2(z) &\sim e^{-\lambda z} z^{E+\lambda^2-\epsilon}
        \left(1+\scO\left(\frac{1}{z}\right) \right),\\
    \end{split}
\end{equation}
as can be checked by direct substitution. Here, $\varrho\leq 1$, so
we conclude that all entire solutions of this equation belong to $\mathscr{H}$. 

With $\epsilon=0$ system~\eqref{eq:syste} has a $\mathbb{Z}_2$
symmetry.  It is invariant with respect to the involution
$\tau:\mathscr{H}^2 \rightarrow \mathscr{H}^2$ given by $ \tau(\psi_1,
\psi_2)(z)= (\psi_2(-z), \psi_1(-z))$. In other words, if $
(\psi_1(z), \psi_2(z))$ is a solution of this system, then also $
(\psi_2(-z), \psi_1(-z))$ is its solution.  We say that a solution
$\psi=(\psi_1, \psi_2)$ of~\eqref{eq:syste} has parity $\sigma\in\{-1,
+1\}$, if $\tau(\psi)=\sigma \psi$.

Analyzing our method of determination of the spectrum for the Rabi
model we noticed several important facts. First of all we asked if the
discrete symmetry of the Rabi problem, whose role was so strongly
underlined in~\cite{Braak:11::}, is really important for determination
of the spectrum. Our answer to this  question is negative.
Amazingly enough, its explicit use in the Rabi model, hides somehow a
good way to attack the problem for which its analytical nature plays
the crucial role. A necessary condition for $\psi(z)$ to be
an eigenvector is that it must be holomorphic in the whole complex
plane. Here it is worth to mention that this is only a necessary
condition, not necessary and sufficient one.  This fact is of crucial
importance for a proper physical interpretation of the obtained
results. Simply, we can mistakenly interpret certain values of energy
as eigenvalues of the Hamiltonian.

\section{Application to the second model}

The general procedure of finding an entire solution is the same for a
system as for a single second order equation. We look for a
holomorphic solution around each singular point and then we ``glue''
them together to one entire solution in that the local solutions
around two singular points must coincide. The technical difference
is that each solution is a vector, so the local series is determined
by a matrix recurrence.

System~\eqref{eq:systmat} has two regular singular points
$s\in\{-1,+1\}$, which are poles of the matrix of coefficients
$\vA(y)$.  We look for local solution around these points that have
the form
\begin{equation}
  F(\rho,y) = (y-s)^{\rho} \sum_{n=0}^{\infty} (y-s)^{n} a_n(\rho),
  \label{eq:form}
\end{equation}
where, $F= \left[ F_1,F_2 \right]^T$, and $a_n= \left[ a_{n,1},a_{n,2}
\right]^T\in\C^{2}$.  To determine $\rho$ and $a_0$ we substitute the
above series into system \eqref{eq:systmat}, and require that the lowest
order term vanishes:
\begin{equation}
  (\vA_s - \rho\id_2)a_0 =0,
\end{equation}
where $\vA_s$ is the residue matrix of $\vA(y)$ at $y=s$. In other
words, $\rho$ must be an eigenvalue of $\vA_s$ and $a_0$ the
corresponding eigenvector. In our case, the poles of $\vA(y)$ are all
simple, so the eigenvalues will be the characteristic exponents.
For both points we simply
have $\rho\in\{0,x\}$, where
\begin{equation}
  x=\frac{4g^2+4\omega E+\omega_0 U}{4\omega^2-U^2},
\end{equation}
and the nonzero exponent is the same spectral parameter as for the
basic Rabi model, i.e. when $U=0$, $\omega=1$ and $g=\lambda$.

Once $a_0$ is determined, the other coefficients are given by a
recurrence relation of the form
\begin{equation}
  \left(\vA_s - (\rho+n)\id_2\right)a_n = R(a_{n-1},\ldots,a_0),
  \label{rec}
\end{equation}
for some linear function $R$.  If $x$ is not an integer, then at both
singular points there are solutions of the form \eqref{eq:form} with
each of the exponents. However, the only solution that can be entire
is the one with $\rho=0$ which we denote by $\psi^{(s,0)}(y):=F(0,y)$.

A local solution around the second singular point can be find in a
similar way. However, we can obtain it in a simpler way. Let us notice
that system~\eqref{eq:systmat} has the following
symmetry. If $\psi(y)$ is its solution, then
\begin{equation}
  \widetilde\psi(y) := \sigma_z\psi(-y),
\end{equation}
is also its solution. Hence
\begin{equation}
  {\widetilde\psi}^{(s,0)}(y) := \sigma_z\psi^{(s,0)}(-y), 
\end{equation}
is a solution of~\eqref{eq:systmat}.  But $\widetilde\psi(y) $ is a
power series in $(y+s)$, so it is a local holomorphic solution around
the other singular point.  In other words
\begin{equation}
  \label{secs}
  \psi^{(-s,0)}(y):=  \sigma_z\psi^{(s,0)}(-y).
\end{equation}
Note that this only gives advantage in calculations because only one
series has to be determined, but the existence of such a symmetry does
not influence the fundamental conditions to be met.

\begin{figure}[ht]
  \begin{center}
    \includegraphics[width=0.9\columnwidth,clip]{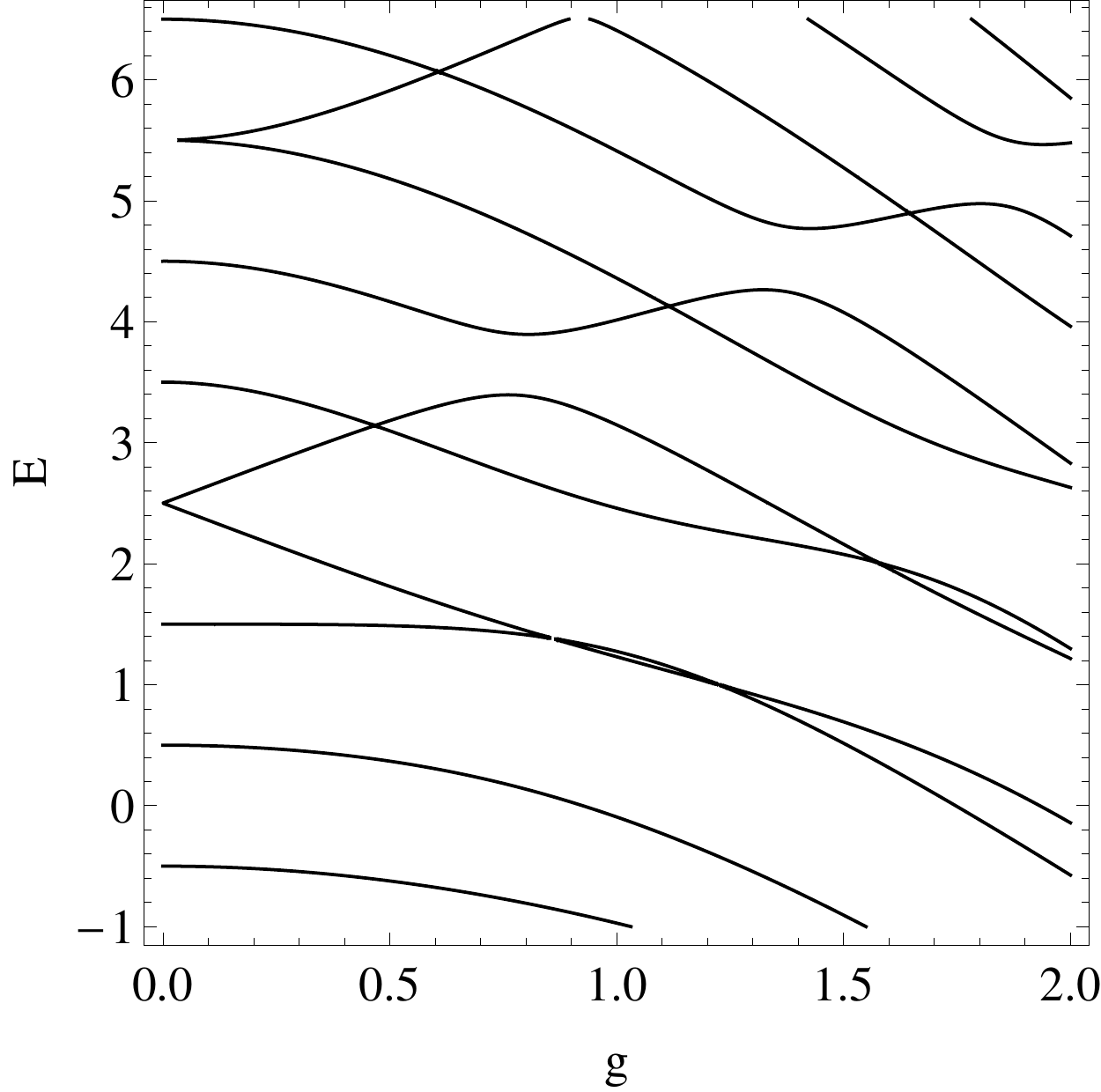}
    \caption{\label{fig:Rabigen_case2} Spectrum of the the second
      generalized Rabi model for $\omega_0=1$, $\omega=2$ and $U=-2$.}
  \end{center}
\end{figure}

If the above two solutions coincide in their common domain
of definition, then their Wronskian
\begin{equation}
  w(p,y) := \det\left[\psi^{(s,0)}(y),\sigma_z\psi^{(s,0)}(-y)\right],
\end{equation}
vanishes for arbitrary $y$. In this model
$p:=(E,\omega,\omega_0,g,U)$.  Exemplary spectrum obtained from
condition $W(p):=w(p,0)=0$ for $\omega=2\omega_0=-U=2$ is shown
in Fig.~\ref{fig:Rabigen_case2}.

When $x$ is a positive integer, say $m$, there might arise logarithmic
terms in the solution, and the form $\eqref{eq:form}$ only works for
one of the exponents.  An elegant way to recover both solutions, due
to Frobenius, is to take the series~\eqref{eq:form} with different, rescaled $a_0(\rho)$ and define solutions as
\begin{equation}
  \begin{split}
    \psi^{(s,x)}(y)&:= F(0,y),\\
    \psi^{(s,l)}(y)&:=
    \left.\frac{\partial F(\rho,y) }{\partial\rho}\right|_{\rho=0}\\
    &= \ln(y-s)F(0,y) + \sum_{n=0}^{\infty} a_n'(0)(y-s)^{n}.
  \end{split}
\end{equation}
The choice of $a_0(\rho)$ must be such that the first
$m-1$ terms of $F$ vanish at $\rho=0$, and the first solution actually
corresponds to the higher integer exponent $x=m$. For a detailed
exposition see \cite{Kohno:99::}.

The determination of the entire solutions here actually unites the
Juddian solutions and the infinite family of non-polynomial ones
discovered in \cite{Maciejewski:14::}.  To wit, the series
$\psi^{(s,x)}$ is always locally analytic, so that it enters into the
usual Wronskian condition of connecting solutions around different
points. If this Wronskian vanishes, then there exists one entire
solution, regardless of the presence of the logarithmic term.

By construction, the second solution $\psi^{(s,l)}$ is such that the
second series is well defined for all values of parameters, so any
condition that the logarithmic term vanishes must be given by a common
factor $J_m$ of all the $a_n$ entering the first series
$F(0,y)$. In such a case, the corresponding coefficient
of $\psi^{(-s,x)}(y)$, which is $\sigma_z a_{m}$, will also contain
that factor. The Wronskian condition will be
\begin{equation}
  \det\left[\psi^{(s,x)}(y),\sigma_z\psi^{(s,x)}(-y)\right]=0,
\end{equation}
and it will include the condition $J_m=0$ because both vectors are
proportional to it. At the same time it will contain the condition for
the solution to be entire in case the logarithm does not vanish
because the two series with exponent $x=m$ around different points
will coincide and define an entire function.

The solutions with $J_m=0$ are the counterparts of the classical
Juddian solutions, and they appear when the parameters lie on some
algebraic curves in the parameter space. Recall that the standard
Juddian curves were $m$ ovals, restricted to a finite region of $g$
and $\omega_0$ and also that for $\omega_0=0$ the system decoupled
trivially.  Here, each $J_m$ has a factor which corresponds to a
deformation of the $\omega_0=0$ case, given by the following parabola
in the $(\omega_0,g)$ plane
\begin{equation}
  \omega_0 = \frac{4g^2U}{4\omega^2-U^2}-Ux.
  \label{judd_para}
\end{equation}
This can be verified by direct substitution into the $\vA$ matrix,
which becomes diagonal, so that the system decouples and can be solved
explicitly
\begin{equation}
  \psi_{1,2} = \exp\left(\mp\frac{4g^2}{4\omega^2-U^2}y\right)(y\pm 1)^x.
\end{equation}
Remarkably \eqref{judd_para} is the only Juddian condition that can be
given explicitly for any integer $x$, in contrast to the other
conditions, which have to be determined recursively. Obviously it is not
confined like the usual ovals and gives nontrivial parameter values
for arbitrarily large $g$. The other conditions are much more involved
than in the classical model, e.g., the first one corresponding to
$x=m=1$ is
\begin{equation}
  \begin{split}
    J_1 = & \left(4g^2U+(U^2-4\omega^2)(U+\omega_0)\right)\\
    & \left(16g^4U^2+(U^2-4\omega^2)^2
      ((U+\omega_0)^2-4\omega^2)+\right.\\
    & \left. 8g^2(U^2-4\omega^2)(U(U+\omega_0)-8\omega^2)\right),
  \end{split}
\end{equation}
where the first factor is the aforementioned parabola. Note that when
the parameters are chosen so that $J_m=0$ the series $F(0,y)$
vanishes, so the actual solution with the higher exponent
is $\tilde{F}$ such that $F(0,y)=J_m\tilde{F}$.

Similarly to the previous model, thanks to the additional coupling,
it is possible for the closed curves defined by $J_m$ to intersect the
parabola \eqref{judd_para} of some $J_n$ in which case there will be
two integer values of $x$ in the spectrum, which is a novel feature of the model.

Regarding the normalization condition, for the system
\eqref{eq:systmat}, one has the expansions
\begin{equation}
  \psi_{\pm}(z)\sim e^{\pm\sigma z} z^{\upsilon}
  \begin{bmatrix}
    \sqrt{2\omega+U} +\scO\left(\tfrac{1}{z}\right) \\[0.5em]
    \mp \sqrt{2\omega-U} +\scO\left(\tfrac{1}{z}\right)
  \end{bmatrix} ,
\end{equation}
with
\begin{equation}
  \sigma = \frac{2g}{\sqrt{4\omega^2-U^2}},
  \qquad \upsilon = \frac{4g^2+4\omega E+\omega_0 U}{4\omega^2-U^2},
\end{equation}
so that the growth order is one, and all entire solutions belong to
$\mathscr{H}$.

\section{Acknowledgements}

The authors wish to thank M.~Ku\'s for stimulating discussions.  This
research has been supported by grant No.~DEC-2011/02/A/ST1/00208 of
National Science Centre of Poland.

\end{document}